\documentclass[journal=jctcce,manuscript=article,hyphens,maxauthors=60]{achemso}
\usepackage{mciteplus}
\usepackage[version=3]{mhchem} 
\usepackage{physics}
\usepackage[T1]{fontenc} 
\usepackage[utf8x]{inputenc}

\usepackage{caption}
\usepackage{soul}
\usepackage{xcolor}
\usepackage{verbatim}
\usepackage{tabularx}
\usepackage{siunitx}
\usepackage{multirow}

\usepackage{listings}
\usepackage{color, colortbl}
\usepackage{lipsum}
\usepackage{subcaption}
\usepackage{graphicx}
\usepackage{booktabs}  
\usepackage{siunitx}

\newcommand{\Lower}[1]{\smash{\lower 1.5ex \hbox{#1}}}
\definecolor{mygreen}{rgb}{0,0.6,0}
\definecolor{mygray}{rgb}{0.5,0.5,0.5}
\definecolor{mymauve}{rgb}{0.58,0,0.82}
\definecolor{python_bg}{RGB}{247, 247, 247}
\definecolor{halfgray}{gray}{0.55}
\definecolor{python_frame}{RGB}{207, 207, 207}

\usepackage[colorinlistoftodos]{todonotes}
\author{Antonina Dobrowolska}
\author{Julian Świerczyński}
\author{Paweł Tecmer}
\affiliation[IF]
{Institute of Physics, Faculty of Physics, Astronomy, and Informatics, Nicolaus Copernicus University in Toruń, Grudziadzka 5, 87-100 Toruń, Poland}
\author{Emil Sujkowski}
\author{Somayeh Ahmadkhani}
\affiliation[IF]
{Institute of Physics, Faculty of Physics, Astronomy, and Informatics, Nicolaus Copernicus University in Toruń, Grudziadzka 5, 87-100 Toruń, Poland}
\altaffiliation
{Present address: Department of Mathematics and Computer Science, Freie Universität in Berlin, Germany}
\author{Grzegorz Mazur}
\affiliation[UJ]
{Department of Computational Methods in Chemistry, Jagiellonian University, Faculty of Chemistry, Jagiellonian University, Gronostajowa 2, 30-387 Kraków, Poland}
\author{Klemens Noga}
\affiliation[CYFRONET]
{Academic Computer Centre Cyfronet AGH, Nawojki 11a, 30-950 Kraków, Poland}
\author{Jeff Hammond}
\affiliation[NVIDIA]
{NVIDIA Helsinki Oy, 00180 Helsinki, Finland}
\author{Katharina Boguslawski}
\email{k.boguslawski@umk.pl}
\affiliation[IF]
{Institute of Physics, Faculty of Physics, Astronomy, and Informatics, Nicolaus Copernicus University in Toruń, Grudziadzka 5, 87-100 Toruń, Poland}

\title[Efficient Coupled-Cluster Python Frameworks]
  {Efficient Coupled-Cluster Python Frameworks for Next-Generation  GPUs: A Comparative Study of CuPy and PyTorch on the Hopper and Grace Hopper Architecture}

\abbreviations{CCSD, GPU, CPU, CuPy, PyTorch, H100, GH200, PyBEST}

\begin{document}

\begin{tocentry}
\includegraphics[width=\columnwidth]{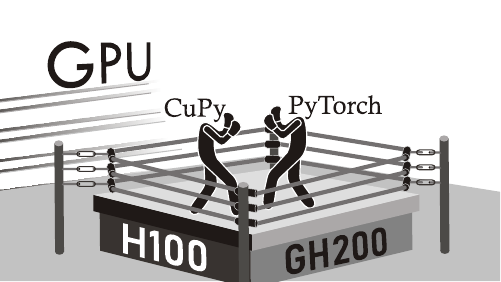}
\end{tocentry}

\begin{abstract}
In this work, we introduce new batching algorithms to effectively handle large contractions encountered in coupled-cluster singles and doubles (CCSD) implementations in Python on the Video Random Access Memory (VRAM) of graphical processing units (GPUs), thereby improving performance.
Specifically, we benchmark the performance of the CuPy and PyTorch libraries on a single NVIDIA Hopper (H100) and the Grace Hopper (GH200) architectures. 
We begin by optimizing the particle-particle ladder bottleneck contraction in CCSD using an asymmetric and dynamic splitting recipe, and then move toward a generic tensor contraction protocol that enables tensor contractions to be performed almost exclusively on GPUs.
We benchmark our new, fully generic GPU-accelerated coupled-cluster implementations for various molecular systems and basis-set sizes, using both the CuPy and PyTorch libraries. 
While PyTorch outperforms CuPy on H100 by approximately 20\%, both perform similarly on the GH200 architecture. 
Compared to our initial GPU implementation [J. Chem. Theory Comput. 2024, 20, 3, 1130–1142], we achieve a 10-fold speedup.
In molecular CCSD calculations, we report additional speedups between 3 and 16 for a single CCSD iteration using Cholesky-decomposed electron repulsion integrals compared to our original GPU-CPU hybrid implementation. 
\end{abstract}
\section{Introduction}\label{sec:introduction}
Rapid advances in modern graphics processing unit (GPU) hardware have fundamentally transformed scientific computing over the past two decades, enabling unprecedented scale and speed in simulations across diverse domains of computer-aided research.
These include, among others, molecular dynamics, climate modeling, drug discovery, astrophysics, materials science, and quantum chemistry, where GPUs facilitate the modeling of complex systems with orders-of-magnitude improvements in throughput compared to traditional central processing unit (CPU)-based approaches.~\cite{gpu-on-summit-covid-proceddings-2020, goetz-nvidia-amd-comparision-jcim-2023, uni-doc-gpu-acceleration-jctc-2023, gpu-accelerated-solar-potential-appl-energy-2024, jax-cpfem-gpu-npj-comput-mat-2025, gpu-accelerated-astrophysics-aa-2025}
At the core of this revolution is the GPU's massively parallel architecture, which efficiently processes large volumes of data by executing independent tasks concurrently on thousands of cores.
The GPU architecture differs fundamentally from that of CPUs, which typically feature only tens of cores and are optimized for sequential processing and complex control-flow workloads.
Another key difference lies in the memory hierarchy.
CPUs rely on general-purpose system RAM (Random Access Memory), which prioritizes low latency and versatility across a wide range of tasks.
GPUs, in contrast, employ dedicated high-bandwidth VRAM (Video Random Access Memory), engineered for rapid, high-throughput parallel access in graphics rendering and compute-intensive workloads.
As a result, when algorithms exhibit high data parallelism, GPUs can deliver speedups of 10 to 100 times over multi-core CPUs, dramatically reducing time-to-solution and energy consumption.~\cite{semiempirical-cpu-gpu-jctc-2012,gpu-4-pyscf-jcpa-2025, byteqc-gpu-wires-2025, pybertha-gpu-jctc-2025}

In quantum chemistry, GPUs excel at computing two-electron repulsion integrals (ERIs),~\cite{gpu-eri-yasuda-jcc-2008, gpu-eri-martinez-jctc-2008, eri-gpu-matinez-jctc-2011, boys-gpu-mazur-jmc-2016} and effective core potentials,~\cite{gpu-ecp-jcp-2015} performing density functional theory (DFT)~\cite{dft-on-gpu-jcp-2009, terachem-wires-2021} and   localized active-space self-consistent field (LASSCF) calculations,~\cite{lasscf-gpu-jctc-2025} and efficiently constructing the Fock matrix---a major computational bottleneck in self-consistent field (SCF) procedures.~\cite{gpu-scf-martinez-jctc-2009, gpu-hf-gordon-jctc-2012, gpu-fock-matrix-appl-sci-2025, gpu-review-goetz-chapter-2010,terachem-wires-2021, gpu-4-pyscf-jcpa-2025}
However, the use of GPUs for large-scale correlated post-Hartree–Fock electronic structure calculations---such as those based on the coupled-cluster (CC) ansatz---remains limited.
Although the limited memory capacity of a single GPU and irregular memory access patterns pose significant challenges to effective scaling for High-Performance Computing (HPC), GPU-accelerated CC implementations have been reported in the literature.~\cite{gpu-ccd-deprince-jctc-2011, non-iterative-ccsdt-kowalski-jctc-2011, gpu-cc-gordon-jctc-2013, density-fitted-ccsd-on-gpu-deprince-mp-2014, ccsd-gpu-v100S-jctc-2020, gpu-cc-dirac-jctc-2021, rank-redued-ccsd-gpu-martinez-jcp-2022, pybest-gpu-jctc-2024, NWChemRecentDevelopments, gpu-ccsdt-jcpa-2026} 
Yet, harnessing the full potential of GPUs may require a serious rewrite of complex electronic structure codes.
Hence, modular, library-heavy electronic structure implementations are desirable, where GPU-accelerated computing is accessible through libraries, and computations on CPUs and GPUs are dynamically accessible via interfaces.\cite{dmrg-h100-gpu-jctc-2024, pybest-gpu-jctc-2024, qc-software-roadmap-electr-structure-2024, qc-software-nat-phys-rev-2025}

As an example, Python offers several open-source libraries~\cite{tensorflow-conf-proc-2016,CuPy-paper-nips-2017,PyTorch-book-chapter-2019} for both CPU~\cite{numpy-paper-nature-2020} and GPU-accelerated computing.~\cite{pybest-gpu-jctc-2024} 
To this end, the rapid development of GPU hardware architectures has led to an immediate growth in Python-based software packages that enable convenient migration of CPU-oriented code to GPUs without requiring low-level programming, making it highly compatible with the NumPy library.~\cite{numpy-paper-nature-2020} 
Examples are CuPy,~\cite{CuPy-paper-nips-2017} PyTorch,~\cite{PyTorch-book-chapter-2019} and TensorFlow.~\cite{tensorflow-conf-proc-2016} 
Although the main driving force behind the development of GPU-supported Python libraries is machine learning, these libraries become practical in quantum chemistry-oriented problems.~\cite{pybest-gpu-jctc-2024, pybertha-gpu-jctc-2025, pyemb-gpu-jcp-2024, byteqc-gpu-wires-2025, pyseqm2.0-jctc-2025}
Recently, we explored a Python-based coupled cluster singles and doubles (CCSD) implementation using Cholesky-decomposed~\cite{cholesky-review-2011} two-electron integrals in the PyBEST software package~\cite{pybest-paper-cpc-2021, pybest-paper-update1-cpc-2024} using the CuPy library.~\cite{pybest-gpu-jctc-2024}
We showed that to work within the 32 GB of VRAM on the NVIDIA Tesla V100S PCIe GPU, tensor contractions must be executed in a batch-wise manner, dividing large intermediate tensors into smaller subproblems that fit available device VRAM.
This batching strategy was carefully tuned for the V100S architecture to maximize efficiency.
Despite this memory constraint, offloading the key tensor contractions to the GPU using CuPy yielded a speedup of 10–16 times for the dominant bottleneck contraction relative to an equivalent NumPy-based implementation running on 36 CPU cores.~\cite{pybest-gpu-jctc-2024}

A solution to the limited VRAM problem on single GPUs, such as the V100S, lies in newer architectures like NVIDIA's Hopper-based systems~\cite{dmrg-h100-gpu-jctc-2024} and the Grace Hopper Superchip (GH200), which additionally offer substantially increased high-bandwidth memory (HBM).~\cite{gh200-article-2024}
A single Hopper GPU in the GH200 provides up to 96 GB of HBM3.
Beyond expanded dedicated GPU memory, the GH200 is a true heterogeneous superchip that tightly integrates an Arm-based Grace CPU (72 cores) with the Hopper GPU via the high-speed NVLink-C2C interconnect.
This coherent, memory-unified design enables low-latency, cache-coherent data sharing between the CPU's system memory and GPU HBM.
As a result, traditional PCIe data-transfer bottlenecks are largely eliminated.
More computational advantage is to be offered within NVIDIA's Blackwell architecture~\cite{blackwell-gpu-webpage},
particularly when exploiting FP64 emulation~\cite{GunnelsFP64emulation,OzakiINT8,Ozaki2FP8},
which has been previously demonstrated for the DGEMM-intensive density-matrix renormalization group (DMRG) method~\cite{LegezaDMRGemulationBlackwell} and the density-matrix purification method for solving the self-consistent field (SCF) equations~\cite{DawsonFP64emulation}.
All together, this motivates us to further develop and optimize Python-based CC implementations specifically for these newer architectures. 
In this paper, we propose and extensively test alternative GPU batching algorithms against the CuPy and PyTorch libraries for various system sizes and contraction types on single H100 and GH200 architectures. 
Most importantly, these new batching approaches can bring us closer to native CUDA implementations for CCSD. 

This work is organized as follows: In Section~\ref{sec:theory}, we discuss the main bottleneck operations in CC calculations. 
Section~\ref{sec:memory} describes algorithmic techniques designed to maximize GPU memory efficiency.
Section~\ref{sec:comput-details} lists computational details. 
Numerical results and the assessment of the GPU to CPU performance are presented in Section~\ref{sec:results}.
Finally, we conclude and provide an outlook in Section~\ref{sec:conclusions}.

\section{Acceleration of CC calculations}\label{sec:theory}

In this work, we build upon our previous GPU-accelerated Pythonic CC implementation.~\cite{pybest-gpu-jctc-2024}
Specifically, we focus on the CC ansatz~\cite{cizek-jcp-1966,cizek-paldus-1971, bartlett-review-arpc-1981, bartlett-review-rmp-2007, helgaker-book-2000,shavitt-bartlett-book-2009},
\begin{equation}
    \ket{\Psi} = e^{\hat T} \ket{\Phi_0}
\end{equation}
where the cluster operator $\hat T$ is restricted to, at most, double excitations, that is, $\hat T = \hat T_2$ or $\hat T = \hat T_1 + \hat T_2$ ($\ket{\Phi_0}$ is some reference wave function like the Hartree--Fock determinant).
Note, however, that the bottleneck operations of CCSD are due to the $\hat T_2$ excitation operator.
Thus, we will focus our discussion on $\hat T_2$-related terms in the CC working equations.
Furthermore, we will consider a spin-free representation, where the amplitudes bear no information about spin degrees of freedom and the CC equations are spin-summed.
The spin-free double excitation operator takes on the form 
\begin{equation}
    \hat T_2 = \frac{1}{2} \sum_{ij}^{\rm o}\sum_{ab}^{\rm v} t_{ij}^{ab} \hat E_{ai} \hat E_{bj}
\end{equation}
with the CCD amplitude $ t_{ij}^{ab}$ and $\hat E_{ai}$ being the singlet excitation operator, running over all occupied (o) and virtual (v) orbitals for the chosen reference determinant $\ket{\Phi_0}$,
\begin{equation}
    \hat E_{ai} = \hat a^\dagger \hat i + \hat {\bar a}^\dagger \hat {\bar i},
\end{equation}
where {$\hat a^\dagger$ ($\hat{\bar a}^\dagger$)} labels electron creation operators for $\alpha$ ($\beta$) electrons, while {$\hat i$ ($\hat{\bar i}$)} indicates the corresponding annihilation operators.

The scaling-determining step in the CCD amplitude equations is the so-called particle-particle ladder term,
\begin{equation}\label{eq:ppl-term}
    0 = \ldots + \sum_{cd} \bra{ab}\ket{cd} t_{ij}^{cd} + \ldots
\end{equation}
where a summation over the indices $i,j,a,b$ is implied and $\bra{ab}\ket{cd}$ are the electron-repulsion integrals in Physicist's notation.
This term determines the well-known formal scaling of the CCD (or CCSD) equations of $\mathcal{O}(o^2v^4)$.
In large-scale modeling, however, full storage of the electron-repulsion integrals (ERI) is typically prohibitive.
As a remedy to save storage, approximate ERIs are used.
One possibility is to exploit Cholesky decomposition of the ERI,~\cite{cholesky-koch-jcp-2003, cholesky-review-2011}
\begin{equation}\label{eq:cholesky}
    \bra{ab}\ket{cd} \approx \sum_{x} L_{ac}^x L_{bd}^x,
\end{equation}
where $x$ indicates the summation over the elements of the Cholesky vectors.
The dimension of $x$ depends on the chosen threshold of the Cholesky decomposition.
For decent to tight thresholds (around $10^{-5}$), we typically have $x \approx 5(o+v)$.
Since we work with real restricted orbitals that result in 8-fold permutational symmetry of the ERI, both Cholesky vectors $L_{ac}^x$ and $L_{bd}^x$ are identical and only one vector needs to be stored.
Using the approximate Cholesky decomposition of the ERI in eq.~\eqref{eq:cholesky}, we can rewrite the CCD (or CCSD) amplitude equations in eq.~\eqref{eq:ppl-term} so that the $n$-th iteration of the CC vector function $^{(n)}\tilde{t}_{ij}^{ab}$ evaluation becomes (again focusing on the particle-particle ladder term only)
\begin{equation}\label{eq:vfunction-chol}
    ^{(n)}\tilde{t}_{ij}^{ab} = \ldots + \sum_{xcd} L_{ac}^x L_{bd}^x t_{ij}^{cd} + \ldots
\end{equation}
A na{\"i}ve summation over the Cholesky vectors may give the impression that the computational complexity increases to {$\mathcal{O}(xo^2v^4)$}.
By defining suitable intermediates, the formal scaling can be, however, reduced to $\mathcal{O}(xo^2v^3)$ (if $x > o^2$, creating an intermediate array of size $xo^2v^2$) or $\mathcal{O}(o^2v^4)$ (if $v > o^2$, creating an intermediate array of size $v^4$), respectively.
In our previous work,~\cite{pybest-gpu-jctc-2024} we GPU-accelerated the evaluation of the Cholesky-decomposed particle-particle ladder term using the latter path, formally generating an intermediate of size $v^4$.
In order to store all intermediates on the VRAM, we proposed a batching recipe in which we split the axes $x$, $a$, $b$, and, if needed, $i$, and perform tensor contraction operations for the subblocks in question.
Overall, by offloading the bottleneck contraction of the CCD/CCSD vector function evaluation, we achieved a speed-up factor of 3-4 within the CuPy library compared to the CPU-only implementation (in Python using various NumPy methods, see also Ref.~\citenum{pybest-gpu-jctc-2024} for details).

In the following, we will use the \texttt{numpy.einsum} subscript convention to label tensor contraction operations.
For instance, the term on the right-hand side of eq.~\eqref{eq:vfunction-chol} can be translated as the \texttt{`xac,xbd,icjd->iajb'} subscript.
Furthermore, for reasons of generality, we won't discriminate the Cholesky-decomposed ERI from their dense representation when dealing with the formal subscript notation of the underlying tensor contraction operation.
Hence, the terms on the right-hand side of eqs.~\eqref{eq:ppl-term} and \eqref{eq:vfunction-chol} will be labeled using the \texttt{`abcd,icjd->iajb'} subscript or, using consecutive letters of the alphabet, the \texttt{`abcd,ecfd->eafb'} subscript, respectively.
For reasons of generality, we will employ the latter notation throughout this work, that is, subscripts in alphabetic order \texttt{`abcd,ecfd->eafb'}.
We should note that, in our notation convention, the first four indices of the \texttt{numpy.einsum} subscript are reserved for the ERI.
Furthermore, these ERI indices need to be expanded in case of Cholesky-decomposed ERI going from \texttt{`abcd'} to \texttt{`xac,xbd'}.

\section{Memory Management}\label{sec:memory}

\begin{figure}[t] 
    \centering
    \includegraphics[width=1.0\textwidth]{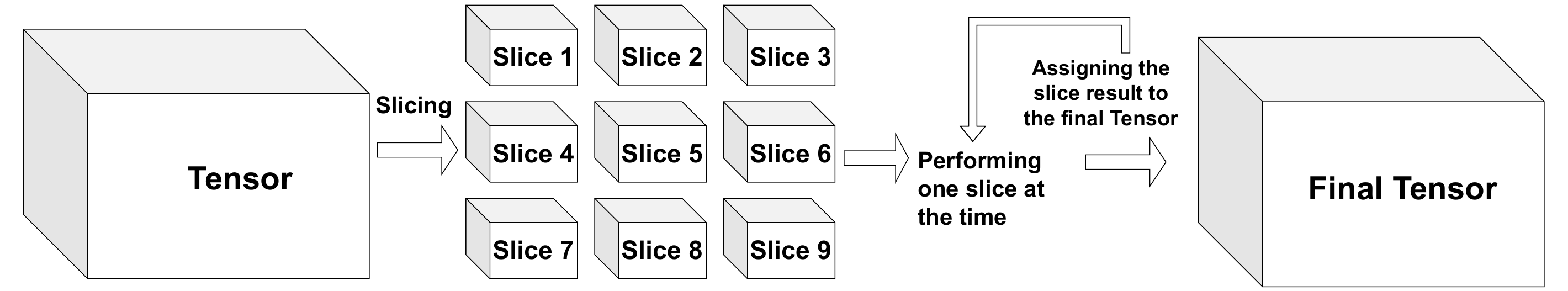} 
    \caption{Schematic batching procedure required to process large tensors when going from the CPU to the GPU and back.}\label{fig:batching}
\end{figure}

As system sizes and/or atomic basis sets increase, the GPU memory becomes a limiting factor.
In contrast to the CPU side, storing the full multidimensional array on the VRAM is impossible with current hardware.
As a remedy, large tensors are split into smaller pieces, then these fragments are moved to the GPU side, where the tensor contraction operations are performed, followed by clearing the GPU memory and then moving the temporary results to the CPU side (see also Figure~\ref{fig:batching} for a schematic representation).
Such batching protocols allow us to handle larger system sizes when the available amount of CUDA memory is smaller than the formally required amount.
In the following, we will scrutinize two different batching protocols.
First, we discuss our optimized batching recipe, which further accelerates the bottleneck operation of CCD/CCSD calculations compared to our previous work.~\cite{pybest-gpu-jctc-2024}
Then, we will introduce a generic batching scheme for arbitrary tensor contractions that are encountered in CCD/CCSD calculations and that support both dense and Cholesky-decomposed ERI, yielding a Pythonic CC implementation that almost exclusively runs on the GPU.

\subsection{An Asymmetric and Dynamic Splitting Protocol}
\begin{figure}[t] 
    \centering
    \includegraphics[width=1.0\textwidth]{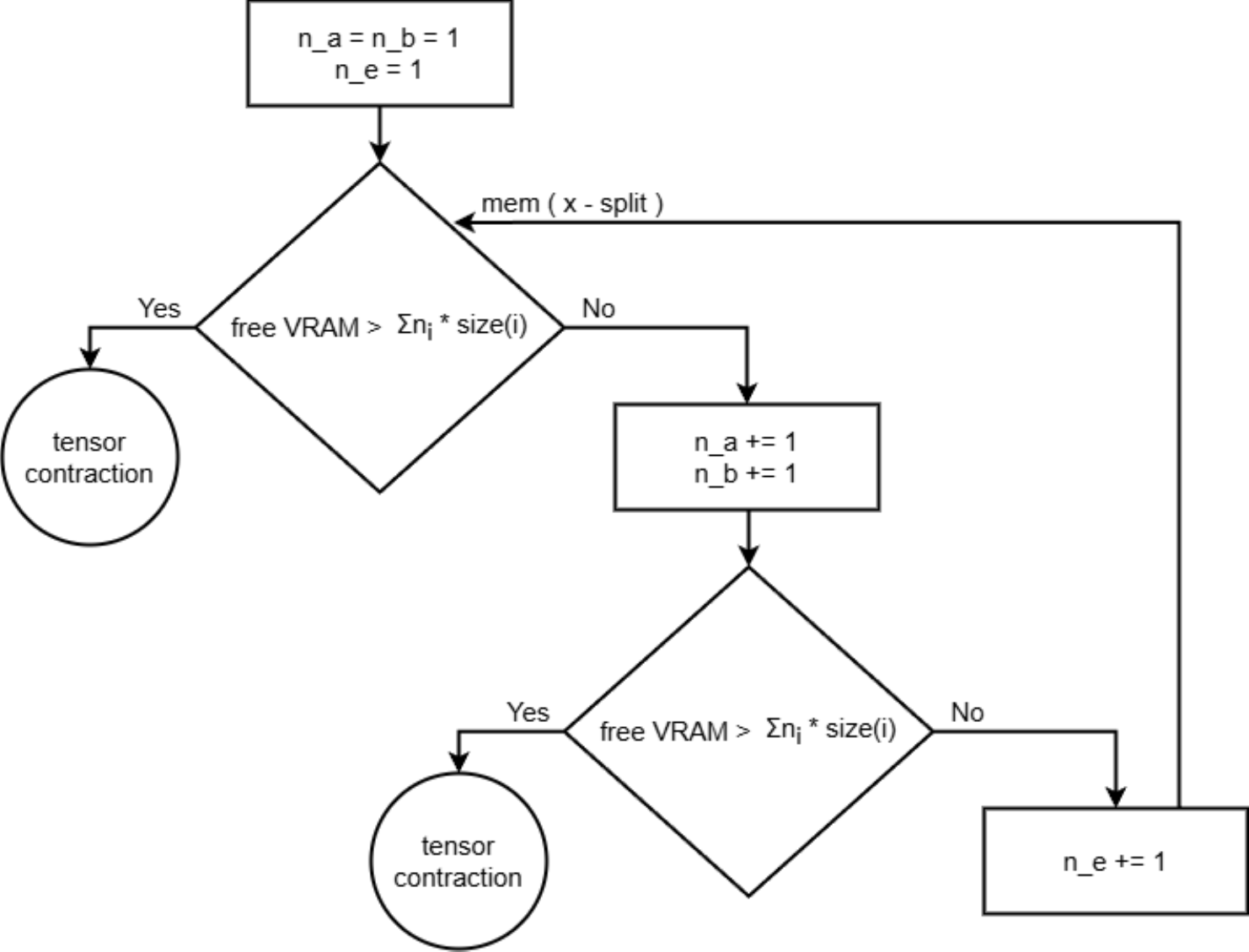} 
    \caption{Schematic illustration of the X-split protocol, where the number of batches is determined along the `\texttt{a}’, `\texttt{b}’, and `\texttt{e}’ axes.
    `\texttt{n\_a}', `\texttt{n\_b}', and `\texttt{n\_e}' indicate the number of batches along the axes `\texttt{a}', `\texttt{b}' and `\texttt{e}', respectively.
    \textit{mem(x-split)} is the memory compute function that determines the VRAM required for the selected tensor contraction operation and is defined in Table~\ref{tbl:memory} and indicated as $\sum n_i * size(i)$ in the decision symbol.
    The splitting process for the Cholesky vectors has been omitted to facilitate a direct comparison with the C-split algorithm.
    }
    \label{fig:pybest_x-split}
\end{figure}

To perform tensor contraction operations involving ``gigantic'' arrays using the GPU, we have to copy all data involved in the contraction scheme to the VRAM in chunks.
In our previous implementation,~\cite{pybest-gpu-jctc-2024} we exclusively focused on the CCD/CCSD bottleneck contraction discussed above.
We should stress again that this contraction is generally labeled as \texttt{`abcd,ecfd->eafb'}, where the first four letters indicate the Cholesky-decomposed ERI, which can be further translated into \texttt{`xac,xbd'}.
Our initial approach comprised an at most three-fold split in the array axes `\texttt{x}’, `\texttt{a}’, `\texttt{b}’ (input arrays), and `\texttt{e}’ (output arrays), while the axes `\texttt{a}’ and `\texttt{b}’ were split homogeneously (that is, in the same number of chunks).
A schematic representation, depicting the decision-making process of the number of chunks along each axis in question, is shown in Figure~\ref{fig:pybest_x-split}.

\begin{figure}[t] 
    \centering
    \caption{Schematic illustration of the C-split protocol, where the number of batches is determined along the `\texttt{a}’, `\texttt{b}’, and `\texttt{c}’ axes.
    `\texttt{n\_a}', `\texttt{n\_b}', and `\texttt{n\_c}' indicate the number of batches along the axes `\texttt{a}', `\texttt{b}' and `\texttt{c}', respectively.
    mem[step0] and mem[step1] are the memory compute function that determine the VRAM required for the selected tensor contraction operation, while free VRAM denotes the accessible VRAM for the current tensor contraction operation.
    All memory functions are defined in Table~\ref{tbl:memory}.
    }\label{fig:pybest_c-split}
    \includegraphics[width=1.0\textwidth]{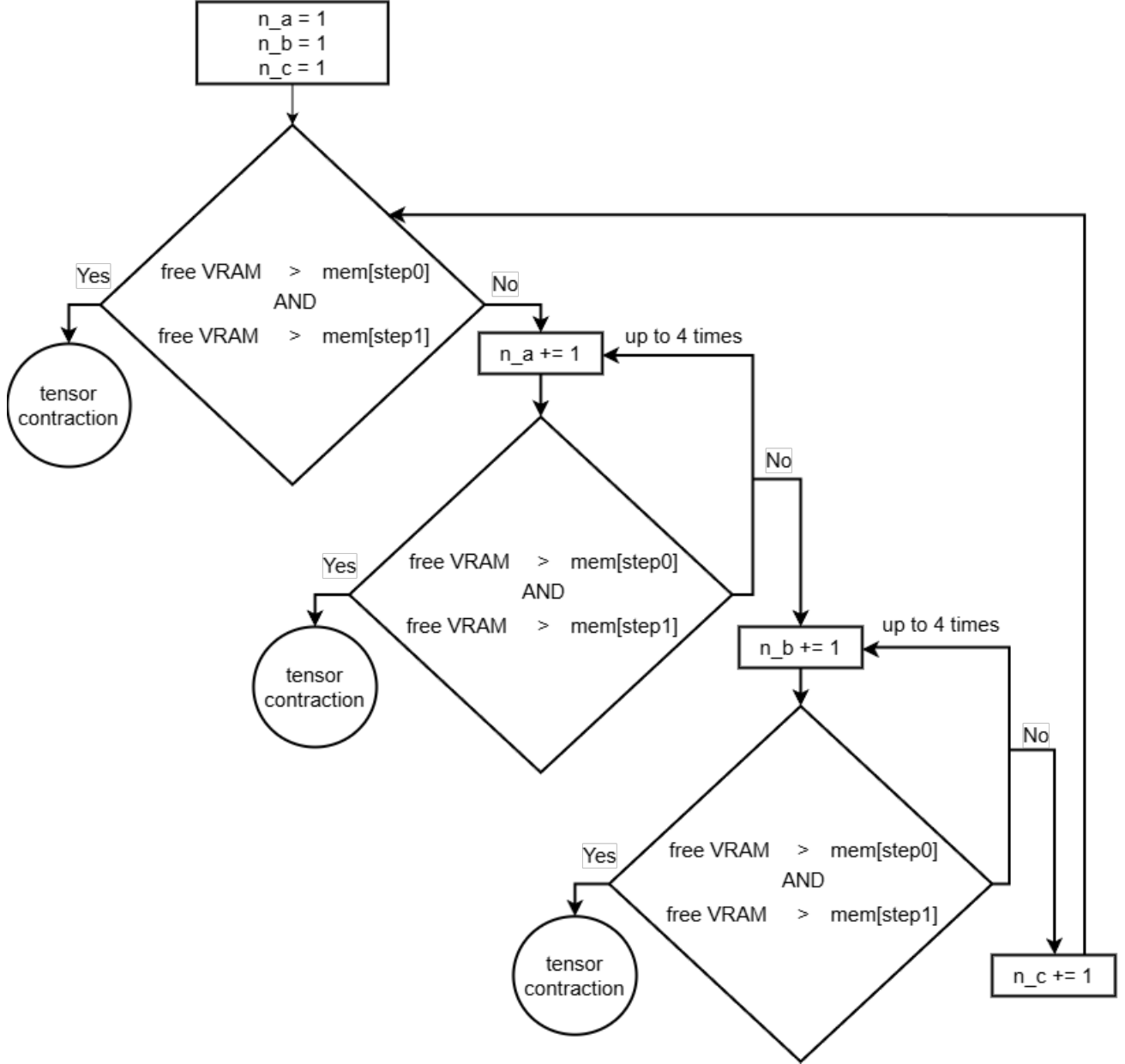}
\end{figure}

 \begin{table}[tb]
\centering
\caption{Memory selection rules to assess the required VRAM compared to the available VRAM for the asymmetric and dynamic batching protocol and its generic extension.
As an example contraction, we choose the \texttt{`abcd,ecfd->efab'} (we further denote this expression as \texttt{op0,op1->out}) tensor contraction for X-split and C-split, where the first `\texttt{abcd}' tensor corresponds to Cholesky-decomposed ERIs and has to be expanded to the `\texttt{xac,xbd}' notation.
For the X-split algorithm, \texttt{n\_chol} is the number of times the \texttt{op0} tensor is split.
Each of the Cholesky-decomposed ERI vectors is divided the same number of times.
The amount of \texttt{op1} batches is determined by \texttt{n\_dense}.
In the C-split batching protocol, the number of splits of the Cholesky-decomposed ERIs is performed with the help of `\texttt{n\_a}', `\texttt{n\_b}' and `\texttt{n\_c}', which correspond to the number of splits along the `\texttt{a}', `\texttt{b}' and `\texttt{c}' axes, respectively.
`\texttt{n\_c}' also determines the amount of \texttt{op1} batches.
For the generic algorithm, the tensor contraction is sequenced in optimal pair contractions, where only the first contraction step is batched according to the notation \texttt{op0,op1->out}.
mem[op]: memory in bytes required for operator op.
n: number of batches for each operator.
The additional prefactor of 2 accounts for intermediates produced by the underlying tensordot operation, as implemented in the GPU library.
Note that for the generic case, op0 and op1 are only batched if the underlying tensors contain indices that are not summed over and thus appear in the output indices.
Since both tensors are, in theory, batched equally, no distinction into n\_0 and n\_1 has been made.
Free VRAM denotes the VRAM reserved for the contraction operations.
Note that for C-split and the generic protocol, the available VRAM (avail.~VRAM) is scaled to reserve space for CuPy/PyTorch processes.
}
\label{tbl:memory}
{\small
\begin{tabular}{l c c c }
\toprule
Splitting & free VRAM & required VRAM  & memory function\\
\hline
\multirow{3}{*}{X-split} &\multirow{3}{*}{avail.~VRAM }& mem[op0] / (n\_chol * n\_chol)\\
& &+ mem[op1] / n\_dense        & mem(x-split)\\
& &+ mem[out] / (n\_chol * n\_chol * n\_dense) \\
\hline
\multirow{6}{*}{C-split} & \multirow{6}{*}{avail.~VRAM * 0.98 * 0.9} & $\max\big($ & $\max\big($\\
& & mem[xac] / (n\_a * n\_c) \\
& & + mem[xbd] / n\_b                       & mem[step0],\\
& &+ 2 * mem[op0] / (n\_a * n\_b * n\_c), \\
& & mem[op0] / (n\_a * n\_b * n\_c) \\
& & + 2 * mem[op1] / n\_c                       & mem[step1]$\big)$\\
& &+ 2 * mem[out] / (n\_a * n\_b)$\big)$\\
\hline
& &  mem[op0] / n \\
generic & avail.~VRAM * 0.9 & + mem[op1] / n    & mem(generic)\\
& & + 2 * mem[out] / n\\
\bottomrule
\end{tabular}
}
\end{table}
Our original implementation was optimized for a specific hardware (NVIDIA V100S), where we screened possible choices in batching for an optimal performance gain.~\cite{pybest-gpu-jctc-2024}
However, as hardware improves, we need to revisit our original partitioning scheme and carefully validate alternative batching recipes.
Current limiting factors are the restriction of the `\texttt{a}’ and `\texttt{b}’ axes to be indistinguishable, resulting them to be split into the same number of chunks, and treating the summation over `\texttt{x}' (first step, \texttt{`xac,xbd->acbd'}) and the summation over `\texttt{c}' and `\texttt{d}' (second step, \texttt{`acbd,ecfd->efab'}) equivalently.
The latter point is particularly crucial as the summations are performed consecutively on the VRAM.
The original logic of deducing batch sizes and performing tensor contractions on them was built as three nested for-loops, which yields cubic scaling in the worst-case scenario (in addition to the formal scaling of the tensor contraction operation in question).
In our new approach, we generalize the batching and consider the consecutive tensor contractions separately when calculating the required memory on the GPU side.
Specifically, we assess whether the tensors in question (as a total or their chunks) would fit on the GPU \textit{after separating} them into the two steps that are calculated sequentially, namely, (i) the first part consisting of \texttt{`xac,xbd->acbd'}, and (ii) the second part being \texttt{`acbd,ecfd->efab'}.
This allows us to adjust the batches more elastically with respect to the size of each tensor.
We also transition from splitting the `\texttt{e}' axis to `\texttt{c}'.
To improve performance for small and medium-sized tensors, we restrict batching of the axes `\texttt{a}' and `\texttt{b}' as long as possible, splitting the `\texttt{c}' axis only when the size of the \texttt{ecfd} tensor is bigger than 40\% of the free and available VRAM on the GPU.
Figure~\ref{fig:pybest_c-split} summarizes a schematic representation of the optimized asymmetric and dynamic batching protocol, while Table~\ref{tbl:memory} collects the selection rules that determine the required VRAM.
We should stress that switching the last batching axis from `\texttt{e}' to `\texttt{c}' was motivated by initial test calculations on the Grace Hopper superchip.
However, we do not present separate benchmark calculations for both choices.
Instead, the `\texttt{e}' axis is split in the original batching recipe, labeled as `X-split', while the optimized batching protocol splits the `\texttt{c}' axis, which is indicated as `C-split'. 

This updated batching protocol contains three very similar, nested for-loops that allow us to batch tensors depending on their unique size.
After the exact number of batches for each axis has been determined, the code snippet below summarizes the procedure of splitting all tensor axes, namely, `\texttt{a}', `\texttt{b}', and `\texttt{c}'.

\begin{lstlisting}[language=Python]
# parts_a, parts_b, parts_c correspond to number of batches we calculated in axes 'a', 'b', 'c' respectively
# Second Cholesky array is split along axis b (x[b]d)
STORE b-batched parts of args[1] as views in cl2
FOR x in parts_b
    APPEND to chol_chunk_lengths_2 size of batched b axis cl2[x]
ENDFOR
# Example: xac,xbd,ecfd -> efab we split:
# [a][b][c]d, e[c]fd -> ef[a][b]
# First dense array args[2] is split along axis c (e[c]fd)
STORE c-batched parts of args[2] as views in operand
# Next, first Cholesky vector is split along axis c (xa[c])
STORE c-batched parts of args[0] as views in cl1

FOR c in parts_c
    MOVE c-th operand view to the GPU and store as operand2 
    # Next, first Cholesky vector is split along axis a (x[a][c])
    MOVE c-th view of a-batched cl1 to the GPU and store as chol_1
    FOR x in parts_a
        APPEND to chol_chunk_lengths_1 size of batched a axis chol_1[x]
    ENDFOR
    SET start_b as the beginning of the view of the part of the matrix we copy to VRAM in axis `b`
    SET end_b as the ending of the view of the part of the matrix we copy to VRAM in axis `b`
    FOR b in parts_b
        ADD chol_chunk_lengths_2[b] to end_b
        SET start_a as the beginning of the view of the part of the matrix we copy to VRAM in axis `a`
        SET end_a as the ending of the view of the part of the matrix we copy to VRAM in axis `a`
        MOVE b-th view of cl2 to the GPU and store as chol_2
        FOR a in parts_a
            ADD chol_chunk_lengths_1[a] to end_a
            COMPUTE tensor contraction using tensordot between asarray_gpu(chol_1[a]) and chol_2 along axes=(0, 0) and store in result_temp
            COMPUTE tensor contraction using tensordot between result_temp and operand2 along axes=([1, 3], [axis_c, axis_d]) and store in result_temp_2

            DELETE result_temp and clean VRAM
            TRANSPOSE result_temp_2 if need and store as result_part
      
            DELETE result_temp_2 and clean VRAM
            
            MOVE result_part to CPU and add to view of final result array defined by start_a, start_b, end_a and end_b
           
            DELETE result_part and clean VRAM
            SET start_a as end_a
        ENDFOR
        SET start_b as end_b
    ENDFOR
ENDFOR
DELETE cl1, cl2, operand2 and clean VRAM
\end{lstlisting}

\subsection{A Generic Batching Recipe}

\begin{figure}[b] 
    \centering
    \caption{Schematic illustration of the generic batching recipe.
    The tensor contraction is sequenced in optimal pair contractions (the ``find optimal path'' decision process), where only the first contraction step is batched according to the notation \texttt{op0,op1->out}.
    `\texttt{n\_0}' and `\texttt{n\_1}' indicate the number of batches along the automatically selected axes of `\texttt{op0}' and `\texttt{op1}', respectively.
    These axes are not summed over and appear both in the input and output arrays, thus ensuring that the batching procedure is passed on to subsequent steps of the optimal contraction path.
    mem(generic) is the memory compute function that determines the VRAM required for the selected tensor contraction operation (labeled as ``ref.'' in the decision process), while free VRAM denotes the accessible VRAM for the current tensor contraction operation.
    All memory functions are defined in Table~\ref{tbl:memory}.
    }\label{fig:pybest_generic-batching}
    \includegraphics[width=1.0\textwidth]{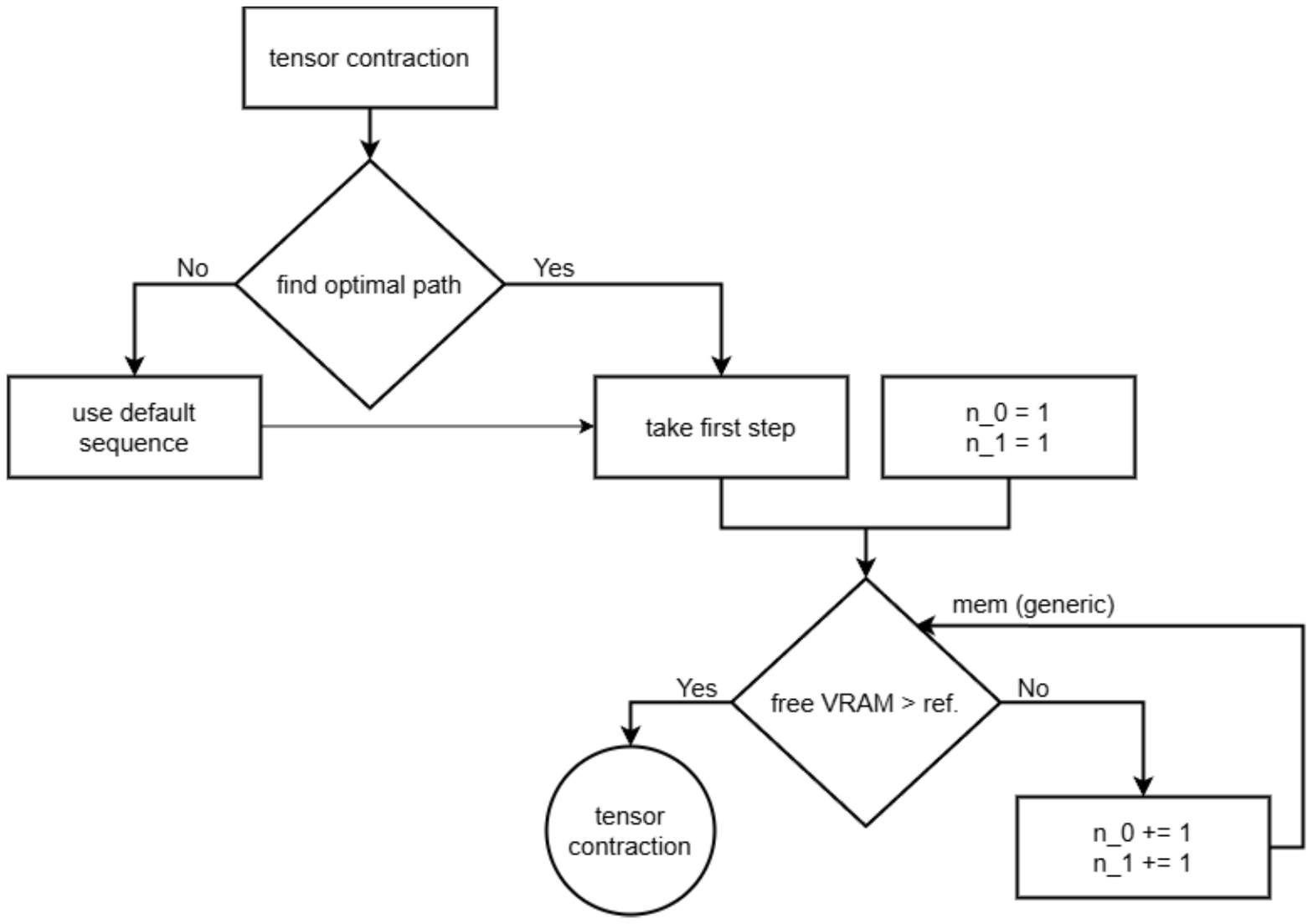}
\end{figure}

Based on our previous work and the improved batching recipe discussed above, we devised a generic batching protocol to offload to the GPU any tensor contractions containing both dense and Cholesky-decomposed arrays.
We should stress that implementing such a generic batching protocol optimally is notoriously difficult and may be best tackled using machine-learning techniques.
In this work, however, we aim for a first, generalizable yet straightforward batching scheme that will even out the required computing times for the bottleneck contraction discussed above and the remaining ones.
In our original CPU-GPU hybrid algorithm,~\cite{pybest-gpu-jctc-2024} the bottleneck operation in CCD/CCSD calculations was reduced from around 70\% to 15\% of the total time per CCD/CCSD iteration step.
Thus, the bottleneck of the CPU-GPU hybrid implementation shifted to the CPU side that features several ``slow'', but ``cheaper'' (in terms of formal scaling) contractions.

For our generic batching protocol, we assume the following:
(i) any tensor contraction involving more than two arrays (for instance, those containing Cholesky-decomposed ERI) are translated into a sequence of two-array contractions.
In the following, we will use the general notation \texttt{op0,op1->out} to label one step in the sequence of contractions.
(ii) An optimal contraction path, that is, an optimal sequence of two-array contractions at a time, is chosen using \texttt{numpy.einsum\_path}.
Note that \texttt{numpy.einsum\_path} yields a contraction sequence of the lowest cost taking into account the creation of intermediate tensors.
If no optimal path is found, an educated guess is made (taking the array sequence as provided in the contraction subscript).
(iii) In the batching protocol, we focus exclusively on the first step of the optimal contraction path.
For this first contraction step, any tensor contraction of the form \texttt{a...,b...->a..b..} will be batched along those axes that are not summed over (here, \texttt{a} and/or \texttt{b}) and simultaneously appear in the \textit{final} output tensor.
If no axes are found, we default to the 0-th axis (dense arrays) or the 1-st axis (Cholesky-decomposed ERI; to avoid splits in \texttt{x}).
The number of batches is increased symmetrically for both operands (labeled as op0 and op1) until all intermediates fit on the VRAM (see Table~\ref{tbl:memory} for the memory evaluation function).
We should note that all subsequent tensor contraction steps are also performed over the batched (intermediate) arrays, until the final step is performed, where the batched result array is added to the correct view of the final result tensor.
The decision to batch the first step of the contraction sequence was made deliberatly, as it represents the first optimal step along the complete contraction pathway.
The second decision to batch the axes that are not summed over is motivated by our updated asymmetric and dynamic splitting protocol (that is, we avoid splitting the \texttt{x} axis).
For reasons of simplicity, we do not support a generic asymmetric batching algorithm yet.  
A flowchart of the batching protocol is depicted in Figure~\ref{fig:pybest_generic-batching}.

Finally, we should note that we offload tensor contraction operations that contain at least one three-dimensional array.
Thus, tensor contractions containing a $T_1$ vertex are not performed on the GPU if they are to be contracted with another dense array.
On the other hand, all tensor contraction operations containing Cholesky-decomposed ERI are offloaded to the GPU, irrespective of the dimensionality of the second operand.
Our choice was motivated by molecular benchmark calculations, which indicated that no significant speed-up has been gained compared to the original CPU implementation.
However, a detailed analysis of such tensor contractions will be the focus of follow-up work.
The pseudo-code snippet below summarizes the logical steps of the generic batching protocol.
\begin{lstlisting}[language=Python]
# Generic GPU helper function performing tensor contractions with tensordot
# Args:
#   subscripts (str): einstein summation label, e.g.,s 'xac,xbd,defc->abfe'
#   operands (np.ndarray): first and second are arrays of Cholesky or dense type
# Returns:
#   np.ndarray: the final output of a tensor contraction

# Sanity check
CHECK subscripts for proper shape

# If possible split subscript into input and output scripts
GET inscripts and outscript from subscripts

# Deduce optimal path for tensordot
GET path from subscripts and operands using np.einsum_path or return a list of tuples

SET mempool to available VRAM

SET op0 to operand[0] and op1 to operand[1]

# Get number of batches for op0 and op1, the corresponding axes (the ones
# that are split/batched), and the label (ind) of the splitted subscripts
# (abcd...)
SET n_batch_0, axis0, ind0, n_batch_1, axis1, ind1 to batch size, axis, and the label (a, b, ..) to be batched of op0 and op1 

# We will batch the first arrays that appear in einsum_path and in the output
# They correspond to operands[step0[0]] and operands[step0[1]]
SET step to path[1]

# Split first two operands contained in first step of path into
# batches under the condition that input and output share axis indices
TAKE op0 as the step[0]-th element of operands list and split it into n_batch_0 along axis axis0
STORE batched array as a view in op0_batched
TAKE op1 as the (step[1]-1)-th element of operands list and split it into n_batch_1 along axis axis1
STORE batched array as a view in op1_batched

# Loop over batched operators
# start_X and end_X indicate the view of the batched array
FOR op0_ in op0_batched
    # Update view of results array for axis0
    # The batched scripts need to show up in the results, otherwise
    # they are simply ignored and the whole array is taken as view
    SET view for axis0 going from first to last index of the current batch
    
    FOR op1_ in op1_batched
        # Update view of results array for axis1
        SET view for axis1 going from first to last index of the current batch
        # Copy subscripts as they get deleted during the batching process
        COPY scripts into scripts_
        # Create a deep copy of the path. Due to batching, we will
        # execute the same path several times
        # Each path gets deleted AFTER it has been executed
        COPY path into path_
        # Create a shallow copy (not copying data) of operand list
        # We loop through the list for each batched view
        # In each iteration, the operands are popped (deleted) from
        # the list. A copy is needed to restart the loop in each
        # batched step
        SHALLOW COPY the list of operands to operands_

        # Loop over all steps in the contraction path
        For step in path_
            # For the 0-th iteration, we need to take the batched arrays
            # For each subsequent path step, we need to pop the next operand
            # contained in the list of operands.
            # The (copied) list of operands contains only unused arrays, that
            # is arrays that have NOT been contracted yet
            POP op0 from operands[step[0]] or take op0_ in first iteration
            POP op1 from operands[step[1]-1] or take op1_ in first iteration
            # Get first subscripts used for op0 (the first in the list of
            # operands)
            POP script0 from scripts_[step[0]]
            # Get view of new op0 for first batched index (ind0)
            # We need to check if ind0 or ind1 are contained in script0
            # If yes, we need to adjust the view, otherwise the take the
            # whole axis
            UPDATE view_0 for batched index ind0
            IF ind1 is in script0 THEN
                UPDATE view_0 for batched index ind1
            ENDIF
            # Update view of op0 so that dimensions matched
            UPDATE view of op0 to match dimension as op0[tuple(view_0)]
            # Subscripts for second operand
            POP script1 from scripts_[step[1] - 1]
            # Find summation axes used in tensordot notation
            FIND axes that are summed over for op0_ and op1_ using tensordot and store them in axis_
            # Find default outscript of tensordot operation
            # It may differ from the outscript of the overal tensor
            # contraction due to the way tensordot works
            # Required for transposition step below
            UPDATE outscript as tensordot may change the order of the axes
            # Move arrays to GPU
            MOVE batched op0 to GPU and store as op0_gpu
            MOVE batched op1 to GPU and store as op1_gpu
            # Do contraction on GPU (finally)
            COMPUTE tensor contraction using tensordot between op0_gpu and op1_gpu along axes=axis_ and store in outmat
            # Cleanup
            DELETE op0_gpu and op1_gpu and clear VRAM
            # Update the list of operands and subscripts
            # The partially contracted array and its subscripts
            # (here outscript_) are appended to the working copies
            # of the operands and scripts. They will be used in the
            # next contraction step
            APPEND outmat to operands_ for next contraction step contained in path_
            APPEND next contraction recipe to scripts_
        ENDFOR
        # Do transposition if required
        IF output array needs to be transposed THEN
            TRANSPOSE outmat
        ENDIF
        # Add batched contraction result to view of result array
        MOVE outmat to CPU and add to view of final result array
        DELETE outmat and clear VRAM
        ENDFOR
    ENDFOR
ENDFOR
\end{lstlisting}

\subsection{Optimized CUDA Memory Allocation in CuPy and PyTorch}
Allocating and deallocating memory using CUDA APIs is time-consuming.
Thus, exploiting CUDA operations like `\texttt{cudaMalloc}' or `\texttt{cudaFree}' with our batching procedures can be costly, especially because the number of batches or slices determines how many memory accesses are required.
To avoid an unnecessary number of CUDA driver calls, PyTorch and CuPy utilize caching memory allocators for tensors.
These allocators optimize memory allocation: whenever a tensor is deallocated, the allocated memory is not returned to the GPU.
Instead, this block of memory is reserved in the cache.
Then, when we need to allocate another tensor, this cache is checked first, and the caching memory allocator ensures that enough memory is available.
This optimized memory allocation procedure speeds up calculations on the GPU side, {but might not be as efficient as calling `\texttt{cudaMallocAsync}' or `\texttt{cudaFreeAsync}' directly.}


\subsection{A Generic Interface for a Dynamic Switching Between CuPy, PyTorch, and CPU-only Libraries}

\begin{figure}[H] 
    \centering
    \includegraphics[width=1.0\textwidth]{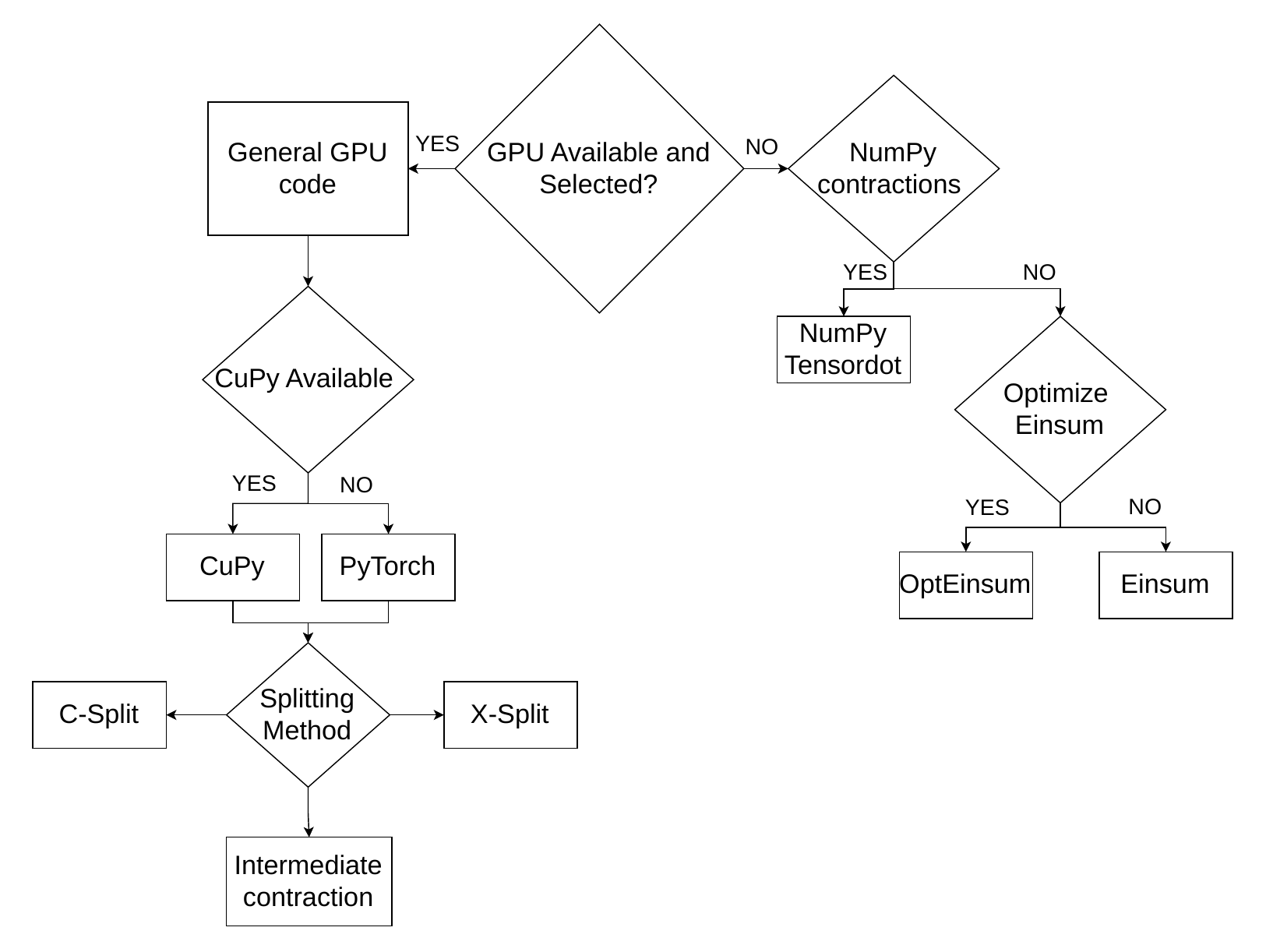} 
    \caption{Schematic representation of the logic flow of the designed tensor contraction engine in PyBEST.
    }\label{fig:pybest_engine}
\end{figure}
Since all CC calculations are dominated by tensor contraction operations, it is desirable to design CC implementations that can exploit any tensor contraction library without changing the codebase.
Based on this idea, PyBEST allows us to define complex array operations concisely using the Einstein summation convention without confining the linear algebra operations to specific library choices.
This allows for a direct comparison of tensor contraction libraries for the same CC codebase.

Considering GPU-acceleration, we always exploit the same batching strategies (discussed above) to optimize the memory handling of tensor contractions on the GPU side.
For ease of use and an optimal flow of logic, we designed and implemented a general GPU-ready tensor contraction engine to change the GPU acceleration background on the fly.
In this work, we will focus on two libraries that provide a Python frontend or Python-based API, namely CuPy and PyTorch.
However, our flexible GPU-ready tensor contraction engine can be easily interfaced with other libraries for GPU-accelerated computing as well (for example, TensorFlow~\cite{tensorflow-conf-proc-2016}).
Currently, our generic tensor contraction function supports tensor operations with a NumPy,~\cite{numpy-paper-nature-2020} CuPy,~\cite{CuPy-paper-nips-2017} PyTorch,~\cite{PyTorch-book-chapter-2019} or C++ backend.

As a least invasive solution, environment variables determine the library choice for GPU acceleration.
Figure~\ref{fig:pybest_engine} shows a flowchart of the implemented GPU-supported tensor contraction logic.
First, we check if a CUDA-ready GPU is available and test basic CUDA operations (allocation and deallocation).
If unavailable or an assertion is raised, we switch to CPU-only tensor contraction libraries.
If implemented and/or supported, \texttt{numpy.tensordot} is taken as the default contraction engine. Otherwise, \texttt{opt\_einsum} or, as a last resort, \texttt{numpy.einsum} is selected (see also Ref.~\citenum{pybest-gpu-jctc-2024} for more details).
At the GPU side, we opt for the CuPy- or PyTorch-based \texttt{tensordot} implementations, while the final output tensors are brought into proper shape using the \texttt{transpose} (CuPy) or \texttt{permute} (PyTorch) functions.
Once the library for GPU acceleration is chosen (CuPy or PyTorch), a batching recipe is selected.
Currently, the three choices described above are supported and cover the `C-split' or `X-split' logic for selected bottleneck contractions or the generic batching protocol. 
This flexible approach yields a modular tensor contraction engine that facilitates testing different libraries for GPU acceleration with minimal changes to the codebase.



\section{Computational Details}\label{sec:comput-details}
All the contraction benchmarks and quantum chemical calculations have been carried out in a developer version of the \texttt{PyBEST}~\cite{pybest-paper-cpc-2021, pybest-paper-update1-cpc-2024} software package (\texttt{v2.2.0-dev0}).
These computations were performed on a single NVIDIA H100 and GH200 architectures available on the LEM computing cluster at the Wrocław Centre for Networking and Supercomputing~\cite{wcss} (WCSS) and the Helios supercomputer cluster at the Academic Computer Centre Cyfronet AGH,~\cite{cyfronet} respectively. 
These computing facilities are part of the Polish Grid Infrastructure (PLGrid).
A detailed description of each GPU/CPU node's hardware and software specifications is provided in Table~\ref{tbl:gpu-spec}. 
\begin{table}[t]
\centering
\caption{Comparison of NVIDIA's GH200 (Helios@Cyfronet) and H100 (Lem@WCSS) hardware and software specifications.}
\label{tbl:gpu-spec}
\begin{tabular}{l S[table-format=5.0] S[table-format=5.0]}
\toprule
\textbf{Parameter} & \textbf{GH200 } & \textbf{H100} \\
\midrule
CPU RAM [GB] & 480 & 1006\\
CUDA Cores & 16896  & 16896 \\
Total VRAM [MiB] & 97871 & 95830 \\
Max Graphics Clock [MHz] & 1980 & 1980 \\
Max Memory Clock [MHz] & 2619 & 1593 \\
Power Limit (TDP) [W] & 900 & 700 \\
CUDA Driver Version & {575.57.08} & {570.195.03} \\
CUDA Toolkit &{12.6.20} & {12.4.99} \\
PCIe Generation (current) & 4 & 5 \\
PCIe Generation (max) & 4 & 5 \\
PCIe Width (current) & 1 & 16 \\
PCIe Width (max) & 1 & 16 \\
Python Version  & {3.11.5} & {3.11.3} \\
NumPy Version & {2.1.3} & {2.3.3} \\
CuPy Version & {13.3.0} & {13.6.0} \\
PyTorch Version & {2.5.1+cu124.post2}  & {2.8.0+cu128} \\
\bottomrule
\end{tabular}
\end{table}

In the contraction benchmark calculations, we tested randomly generated numbers, with timings averaged over five individual runs. 
The first iteration is usually slower because the library must initialize the CUDA driver and its related components. 
To avoid this, we run the GPU jobs with small data allocations first, prior to the main calculations.

\begin{figure}
\centering
\includegraphics[scale=0.95]{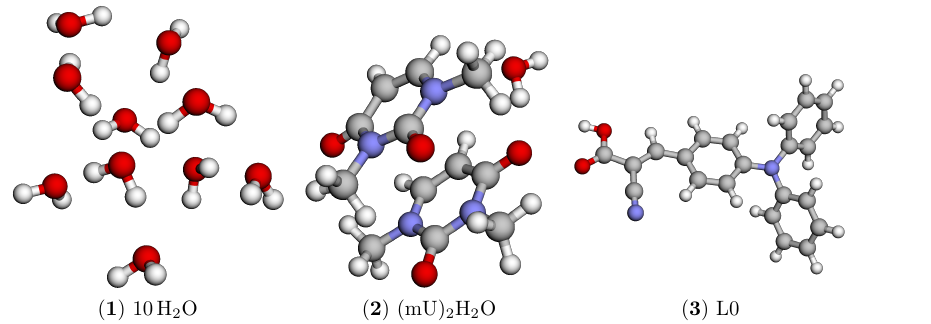}
\caption{Structures of the investigated molecules drawn with the \texttt{PyBEST GUI}. Atom types are color-coded: white for hydrogen, red for oxygen, gray for carbon, and purple for nitrogen. 
}
\label{fig:structures}
\end{figure}

In all molecular calculations, we used the Cholesky-decomposed two-electron integrals with a predefined threshold ($10^{-5}$ unless stated otherwise), and the core orbitals (1s for C, N, and O) were kept frozen.
For the a decameric water cluster---\ce{(H_2O)_{10}} (\textbf{1}) we used the {cc-pVDZ and cc-pVTZ basis sets}.~\cite{cc-pvxz-dunning-jcp-1989}
The 6-31+G** basis set~\cite{6-31G-jcp-1971, 6-31G-jcp-1972, 6-31G*-tca-1973} was used for the a hydrated methylated uracil dimer \ce{(mU)2H2O} complex (\textbf{2}) to compare with existing reference GPU benchmarks.
For the donor-$\pi$-acceptor dye 4-(diphenylamino)phenylcyanoacrylic acidmolecule---L0 dye  (\textbf{3}) we employed the {cc-pVDZ, cc-pVTZ}~\cite{cc-pvxz-dunning-jcp-1989}, and aug-cc-pVDZ basis sets.~\cite{aug-cc-pvxz-jcp-1992}
The xyz structure for all investigated molecules are visualized with \texttt{PyBEST GUI}~\cite{pybest-gui-ijqc-2025} in Figure~\ref{fig:structures}.

\section{Results and Discussion}\label{sec:results}
\subsection{Contraction benchmarks with predefined dimensions and randomly generated arrays}
\begin{figure}
\centering
\includegraphics[width=1\textwidth]{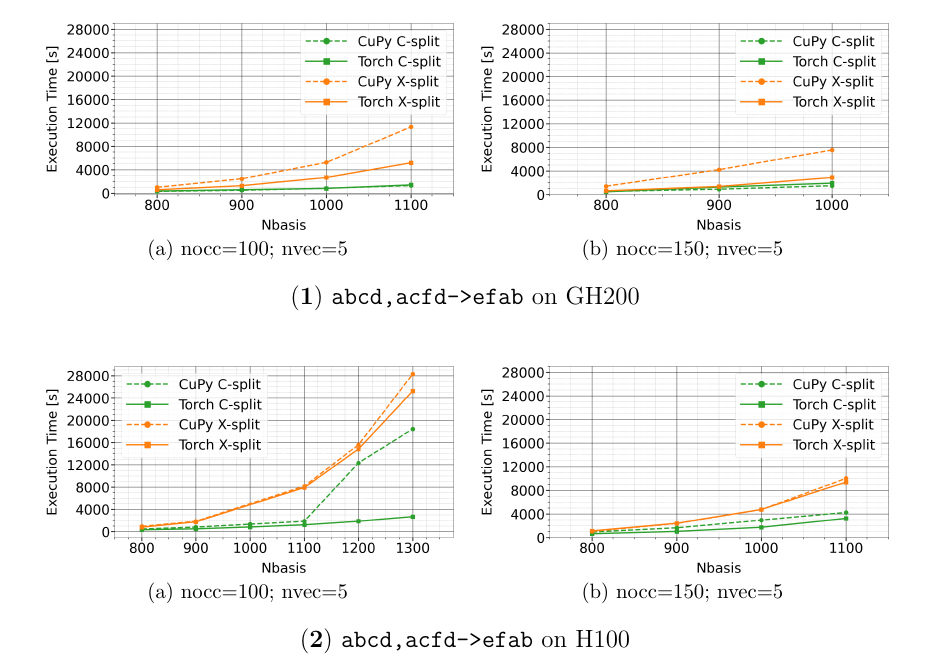}
\caption{Comparison of CuPy and PyTorch-based C and X-splits execution times for the \texttt{abcd,ecfd->efab} contraction (vvvv block) on GH200 and H100 GPU architectures.  
The number of occupied orbitals (nocc) and the number of Cholesky vectors (nvec) in units of $N_{\rm basis}$ is denoted explicitly below each graph.
}
\label{fig:h100-vs-gh200}
\end{figure}

First, we examine GPU performance using controlled synthetic benchmarks with predefined tensor dimensions and randomly generated data (cf.~Section~\ref{sec:comput-details} for details) before proceeding to realistic molecular test cases.
Specifically, we assess the performance of the tensor contraction \texttt{abcd,ecfd->efab} on GPUs across a range of basis set sizes (800–1300 basis functions) and numbers of occupied orbitals (100 and 150).
The benchmarks were carried out using both PyTorch- and CuPy-based implementations of the X-split and C-split algorithms (see Section~\ref{sec:memory} for details on the memory-efficient splitting strategies).
To mimic the computationally dominant contraction step in CCSD for medium-to-large molecules, the number of Cholesky vectors was fixed at $5 \cdot N_{\rm basis}$, corresponding to an approximate decomposition threshold of $10^{-5}$.
Performance results obtained on the NVIDIA GH200 and H100 GPU architectures are shown in the upper and lower panels of Figure~\ref{fig:h100-vs-gh200}, respectively.
The numerical data underlying this figure are provided Table~\ref{tbl:x-c-split-contractions}. 

In all cases examined in Figure~\ref{fig:h100-vs-gh200}, the X-split algorithm---originally optimized for the NVIDIA V100S with 32 GB VRAM---is substantially slower than the newly introduced C-split algorithm.
This holds true across both CuPy and PyTorch implementations and for both the GH200 and H100 GPU architectures.
Both CuPy and PyTorch exhibit markedly better performance on the GH200 compared to the H100, as clearly visible when comparing the upper and lower panels of Figure~\ref{fig:h100-vs-gh200}.
Notably, on the GH200, the combination of CuPy with X-split remains computationally more efficient than PyTorch with X-split, whereas the two libraries perform very similarly when using X-split on the H100.
The performance advantage of C-split over X-split becomes more pronounced as the number of (virtual) basis functions increases (corresponding, in our test suite, to fewer occupied orbitals).
On the GH200 (upper panel of Figure~\ref{fig:h100-vs-gh200}), the difference in performance between CuPy-based and PyTorch-based C-split implementations is negligible.
In contrast, this difference is significantly larger on the H100, particularly for the larger $N_{\rm basis}$ dimension shown in the lower-left part of the figure.
In these cases, the combination of PyTorch with C-split substantially outperforms that of CuPy and C-split, yielding considerably shorter execution times.
In the largest case with $N_{\rm basis}=1300$, the C-split performs 32, 7, 9 slices (corresponding to \texttt{n\_a}, \texttt{n\_b}, and \texttt{n\_c} on Figure~\ref{fig:pybest_c-split}) and X-split executes 20, 19 slices (corresponding to \texttt{n\_a} and \texttt{n\_b} on Figure~\ref{fig:pybest_x-split}), respectively.  

In summary, a significant speed-up in execution times for the \texttt{abcd,ecfd->efab} contraction is obtained by combining PyTorch with the new C-split algorithm. 
The PyTorch-based C-split gives speed-ups of up to a factor of 10 compared to our initial CuPy-based X-split implementation~\cite{pybest-gpu-jctc-2024}.
Furthermore, the use of the NVIDIA GH200 and H100 GPU architectures overcomes the previous limitation on the maximum number of basis functions that can be handled on a single GPU, compared to earlier hardware generations such as the NVIDIA V100S (32 GB).~\cite{pybest-gpu-jctc-2024}
Altogether, this strongly motivates further exploration and application of this approach to realistic molecular test sets.


\begin{table}[t]
 \caption{Comparison of X-split and C-split performance on H100 and GH200 GPU architectures for the \texttt{abcd,ecfd->efab} bottleneck contraction for the particle-particle ladder term using predefined dimensions and randomly generated arrays with CuPy and PyTorch.
 nocc: number of occupied orbitals.
 nvir: number of virtual orbitals.
 nvec: number of Cholesky vectors.
 N: number of basis functions (nocc+nvir).
 n.c.: not computed due to insufficient memory on the CPU side or \texttt{OutOfMemoryError} for X-split. We should note that the memory function for X-split in Table~\ref{tbl:memory} was optimized for medium-sized problems. For reasons of reproducibility and a direct comparison with our original work,~\cite{pybest-gpu-jctc-2024} we did not re-optimized it in this work.
 Note that all GPU times include data preparation (batching on the CPU side), data transfer, and the actual algebraic operations.
 }
\label{tbl:x-c-split-contractions}
\begin{footnotesize}
  \begin{tabular}{ccccc|cc|cc}

  \hline\hline
    \multicolumn{5}{c|}{}&\multicolumn{2}{|c|}{GH200 (avg of 5)} & \multicolumn{2}{|c}{H100 (avg of 5)} \\
    \hline
    \textbf{N}&\textbf{nocc}&\textbf{nvec}&\textbf{nvir}&\textbf{library}&\textbf{X-split [s]}&\textbf{C-split [s]}&\textbf{X-split [s]}&\textbf{C-split [s]}\\
    \hline\hline
    \multirow{2}{*}{800}  & \multirow{2}{*}{200}& \multirow{2}{*}{10 x N} &\multirow{2}{*}{600} &PyTorch &834.7&1573.6&1568.1&1485.3\\ 
    & & & & CuPy & 2061.3 & 706.8 &1068.4&1364.7\\ \hline

     \multirow{2}{*}{900} & \multirow{2}{*}{225}  &\multirow{2}{*}{10 x N}&\multirow{2}{*}{675} &PyTorch &n.c.&n.c.&3628.4&3042.2\\ 
    & & & & CuPy & n.c. & n.c. &13725.0&2529.6\\ \hline

    \multirow{2}{*}{1000}  & \multirow{2}{*}{250} &\multirow{2}{*}{10 x N} &\multirow{2}{*}{750} &PyTorch &n.c.&n.c.&n.c.&6233.1\\ 
    & & & & CuPy & n.c. & n.c. &n.c.&7759.1\\ \hline

    \multirow{2}{*}{800}  & \multirow{2}{*}{100}  &\multirow{2}{*}{5 x N}&\multirow{2}{*}{700} &PyTorch &574.7 &355.9 &789.4&330.5\\ 
    & & & & CuPy &987.7   &306.6   &933.4&461.5\\ \hline

     \multirow{2}{*}{900} & \multirow{2}{*}{100}  &\multirow{2}{*}{5 x N}&\multirow{2}{*}{800} &PyTorch &1265.7&587.1&1753.3&496.2\\ 
    & & & & CuPy & 2450.2 & 495.0 &1853.0&809.7\\ \hline

    \multirow{2}{*}{1000}  & \multirow{2}{*}{100} &\multirow{2}{*}{5 x N}&\multirow{2}{*}{900} &PyTorch &2673.7&805.8&n.c.&827.1\\ 
    & & & & CuPy &5241.7 & 829.1 &n.c.&1338.7\\ \hline

    \multirow{2}{*}{1100}  & \multirow{2}{*}{100} &\multirow{2}{*}{5 x N}&\multirow{2}{*}{1000} &PyTorch &5177.4&1401.7&7907.7&1223.4\\ 
    & & & & CuPy &11305.0 &1282.7 &8151.5&1860.9\\ \hline

    \multirow{2}{*}{1200}  & \multirow{2}{*}{100} &\multirow{2}{*}{5 x N}&\multirow{2}{*}{1100} &PyTorch &n.c.&n.c.&14821.7&1864.7\\ 
    & & & & CuPy &n.c. & n.c. &15601.2&12329.8\\ \hline

    \multirow{2}{*}{1300}  & \multirow{2}{*}{100} &\multirow{2}{*}{5 x N}&\multirow{2}{*}{1200} &PyTorch &n.c.&n.c.&25244.9&2679.1\\ 
    & & & & CuPy &n.c. & n.c. &28292.9&18426.4\\ \hline

    \multirow{2}{*}{800}  & \multirow{2}{*}{150} &\multirow{2}{*}{5 x N}&\multirow{2}{*}{650} &PyTorch &637.4 &475.3 &1128.1&651.2\\ 
    & & & & CuPy &1410.7   &532.8   &1050.2&940.9\\ \hline

     \multirow{2}{*}{900} & \multirow{2}{*}{150} &\multirow{2}{*}{5 x N}&\multirow{2}{*}{750} &PyTorch &1364.7 &1261.1 &2419.8&1044.4\\ 
    & & & & CuPy & 4213.1 & 911.6 &2390.6&1654.1\\ \hline

    \multirow{2}{*}{1000}  & \multirow{2}{*}{150} &\multirow{2}{*}{5 x N}&\multirow{2}{*}{850} &PyTorch &2886.4&1924.4&4749.6&1739.7\\ 
    & & & & CuPy & 7536.1& 1489.2&4727.1&2945.2\\ \hline

    \multirow{2}{*}{1100}  & \multirow{2}{*}{150} &\multirow{2}{*}{5 x N}&\multirow{2}{*}{950} &PyTorch &n.c.&n.c.&9384.7&3234.4\\ 
    & & & & CuPy &n.c.& n.c.&10024.0&4241.5\\ \hline

    \multirow{2}{*}{1000}  & \multirow{2}{*}{50} &\multirow{2}{*}{10 x N}&\multirow{2}{*}{950} &PyTorch &5262.6&574.6&n.c.&673.0\\ 
    & & & &CuPy & n.c.& 763.0&n.c.&1006.4\\ \hline
    
    \multirow{2}{*}{1200}  & \multirow{2}{*}{50} &\multirow{2}{*}{10 x N}&\multirow{2}{*}{1150} &PyTorch &n.c.&1756.0&n.c.&2031.5\\ 
    & & & &CuPy & n.c.& 2155.2&n.c.&2429.5\\ \hline
    \hline\hline
  \end{tabular}
\end{footnotesize}
\end{table}

\subsection{Molecular benchmark calculations.}
Our molecular test set comprises three systems: a decameric water cluster \ce{(H2O)10} labeled as (\textbf{1}), a hydrated methylated uracil dimer, \ce{(mU)2 \cdot H2O} molecule labeled as (\textbf{2}), and the donor-$\pi$-acceptor dye 4-(diphenylamino)phenylcyanoacrylic acid abbreviated as L0 and labeled as (\textbf{3}) in different basis sets (cf. Section~\ref{sec:comput-details}).
Corresponding molecular structures are presented in Figure~\ref{fig:structures}.
These systems are established benchmarks with available reference CC timings from prior GPU implementations. 

The performance of our newly developed CC implementation within PyBEST and C-split is reported and compared with existing references from Psi4 and TeraChem in Table~\ref{tbl:timing2}.
Specifically, we analyze the timing of a single CCSD iteration step and the average time spent on GPU function calls.
Note that not all CCSD calculations are performed exclusively on the GPU unless explicitly stated.
Furthermore, the GPU-accelerated CCSD calculations within PyBEST combine the C-split and generic batching protocols and simultaneously exclude all tensor contractions that cannot be performed using \texttt{tensordot} or contain a dense ERI tensor contracted with a $\hat T_1$ vertex.
All these special cases are still restricted to the CPU.

For the water cluster (\textbf{1}) with the cc-pVDZ basis set, we improved the CCSD timings on GH200 by approximately a factor of three compared to our original CuPy/X-split implementation.~\cite{pybest-gpu-jctc-2024}
For the cc-pVDZ and cc-pVTZ basis sets, the lowest average GPU times are obtained with CuPy on GH200.
On H100, PyTorch is almost 50\% faster than CuPy. 
{Most likely, PyTorch does a better job of hiding PCIe overheads, spending far less wall-clock time waiting for PCIe by overlapping transfers with computation. Still, it is difficult to provide a definitive answer.} 

A similar PyTorch/CuPy performance trend is observed for the uracil dimer monohydrate (\textbf{2}) and the L0 molecule (\textbf{3}) using the cc-pVDZ basis set.
Interestingly, when the number of basis functions increases to over 700 (using the aug-cc-pVDZ basis set), PyTorch provides better timings than CuPy.
However, for the largest investigated system (molecule (\textbf{3}) with the cc-pVTZ basis set and more than 1000 basis functions), only CuPy computations are feasible.
{Importantly, we also see clear evidence that, for this large system, the CPU part of the computations becomes important: CuPy calculations on H100 with 32 CPUs reduce the timing by half compared to a single CPU. 
That suggests the particle particle hole term is no longer the only bottleneck operation. 
}

\begin{table}[t]
\begin{scriptsize}
  \begin{tabular}{llccccc}
  \hline\hline
    \textbf{Basis set}&\textbf{Software}&\textbf{Reference}&\textbf{CPU cores}&\textbf{GPU}&\textbf{CCSD}&\textbf{avg. GPU}\\
    \hline\hline
    \multicolumn{7}{c}{(\ce{ H2O})$_{10}$, 30 atoms}\\
    \hline
    \multirow{8}{*}{\textbf{\shortstack{cc-pVDZ \\ 240 AOs}}}&\texttt{Psi4}/DF&~\citenum{psi4-gpu-jctc-2019}&16&-&16s&-\\
    &\texttt{PyBEST}/CD&~\citenum{pybest-gpu-jctc-2024}&36&-&337s&-\\
    &\texttt{PyBEST}/CD/CuPy&~\citenum{pybest-gpu-jctc-2024}&36&Tesla V100S&92s&4.4s\\
    &\texttt{PyBEST}/CD/PyTorch&this work&1&H100&31.7s&21s\\
    &\texttt{PyBEST}/CD/CuPy&this work&1&H100&48s&37s\\
    &\texttt{PyBEST}/CD/PyTorch&this work&72&GH200&25.4s&17s\\
    &\texttt{PyBEST}/CD/CuPy&this work&72&GH200&23.9s&15.8s\\
    &\texttt{TeraChem}/CD&~\citenum{ccsd-gpu-v100S-jctc-2020}&1&Tesla V100&10s&10s\\
    \hline
    \multirow{4}{*}{\textbf{\shortstack{cc-pVTZ \\ 580 AOs}}}&\texttt{PyBEST}/CD/PyTorch&this work&1&H100&6.4m&4.7m\\
    &\texttt{PyBEST}/CD/CuPy&this work&1&H100&9.6m&7.9m\\
    &\texttt{PyBEST}/CD/PyTorch&this work&72&GH200&5.7m&4.6m\\
    &\texttt{PyBEST}/CD/CuPy&this work&72&GH200&5.5m&4.3m\\
    \hline\hline
    \multicolumn{7}{c}{(mU)\ce{_2 H2O}, 39 atoms}\\
    \hline
    \multirow{7}{*}{\textbf{\shortstack{6-31+G** \\ 468 AOs}}}&\texttt{PyBEST}/CD&~\citenum{pybest-gpu-jctc-2024}&36&-&96.5m&-\\
    &\texttt{PyBEST}/CD/CuPy&~\citenum{pybest-gpu-jctc-2024}&36&Tesla V100S&33.2m&4.3m\\
    &\texttt{PyBEST}/CD/CuPy&this work&1&H100&9.3m&6.8m\\
    &\texttt{PyBEST}/CD/PyTorch&this work&1&H100&6.8m&4.3m\\
    &\texttt{PyBEST}/CD/CuPy&this work&72&GH200&6.1m&4.5m\\
    &\texttt{PyBEST}/CD/PyTorch&this work&72&GH200&6.7m&5m\\
    &\texttt{TeraChem}/CD&~\citenum{ccsd-gpu-v100S-jctc-2020}&1&Tesla V100&2.5m&2.5m\\
    \hline\hline
     \multicolumn{7}{c}{L0, 42 atoms}\\
    \hline
    \multirow{6}{*}{\textbf{\shortstack{cc-pVDZ \\ 444 AOs}}}&\texttt{PyBEST}/CD&~\citenum{pybest-gpu-jctc-2024}&36& -- &45.9m&--\\
    &\texttt{PyBEST}/CD/CuPy&~\citenum{pybest-gpu-jctc-2024}&36&Tesla V100S &15.3m&2.0m\\
    &\texttt{PyBEST}/CD/PyTorch&this work&1&H100&3.2m&2.1m\\
    &\texttt{PyBEST}/CD/CuPy&this work&1&H100&4.5m&3.3m\\
    &\texttt{PyBEST}/CD/PyTorch&this work&72&GH200&3.3m&2.5m\\
    &\texttt{PyBEST}/CD/CuPy&this work&72&GH200&3.1m&2.4m\\
    \hline
    \multirow{4}{*}{\textbf{\shortstack{aug-cc-pVDZ \\ 742 AOs}}}&\texttt{PyBEST}/CD/PyTorch&this work&1&H100&19.5m&15.1m\\
    &\texttt{PyBEST}/CD/CuPy&this work&1&H100&31.2m&26.9m\\
    &\texttt{PyBEST}/CD/PyTorch&this work&72&GH200&17.1m&12.8m\\
    &\texttt{PyBEST}/CD/CuPy&this work&72&GH200&19.7m&15.2m\\
    \hline
    \multirow{6}{*}{\textbf{\shortstack{cc-pVTZ \\ 1004 AOs}}}&\texttt{PyBEST}/CD/PyTorch&this work&1&H100&n.c.&n.c.\\
    &\texttt{PyBEST}/CD/CuPy&this work&1&H100&139.7m&54.6m\\
    &\texttt{PyBEST}/CD/CuPy&this work&32&H100&63.7m&50.8m\\
    &\texttt{PyBEST}/CD/PyTorch&this work&72&GH200&n.c.&n.c.\\
    &\texttt{PyBEST}/CD/CuPy&this work&72&GH200&52.8m&33.6m\\
    &\texttt{PyBEST}/CD/CuPy/GPU-only&this work&72&GH200&59.6m&40.0m\\
    \hline\hline
  \end{tabular}
\end{scriptsize}
  \caption{Average timings [s] of a CCSD iteration step compared to other GPU computations and average GPU time used in one iteration for the investigated molecules (\textbf{1}), (\textbf{2}), and (\textbf{3}) and different basis set sizes.
  One iteration step contains the evaluation of the vector function, the update of the CC amplitudes, and the evaluation of the CC energy expression. All timings correspond to differences in epoch times.
  All PyBEST results are shown for the CPU-only (if available) and various CPU-GPU hybrid variants.
  The \texttt{Psi4} data is given as a comparison.
  CCSD: mean value for the time of one CC iteration averaged over 4 steps.
  avg. GPU: mean value for the time spent in the GPU function call, averaged over 4 steps
  CD: Cholesky decomposition. DF: Density Fitting.
 Note that all GPU times include data preparation (batching on the CPU side), data transfer, and the actual algebraic operations.
  }\label{tbl:timing2}
\end{table}

For the largest basis set in the L0 test case, we also performed computations using a fully generic GPU path, lifting the restriction on contractions involving $\hat T_1$ vertices (note that only \texttt{tensordot}-ready contractions are offloaded to the GPU).
This approach provides similar timings as the restricted GPU variant (excluding all $\hat T_1$ terms from being offloaded to the GPU), albeit the GPU-only variant performs slightly worse.
We should stress here that the GPU-acceleration performance is substantially better on NVIDIA GH200 than H100 for $N_{\rm basis} \ge 1000$.
Furthermore, 30\% of the computing time of the GPU-only approach is still performed on the CPU side.
These operations include data preparation (about 7m; expanding the symmetry-unique $\hat T_2$ amplitudes to their dense representation to allow for numpy manipulations) and other tensor contractions (around 13m) that cannot be performed using the \texttt{tensordot} method.
Examples for the latter are the creation of the $\tilde{t}^{cd}_{kl} = t^c_k t_l^d$ intermediate (during the residual evaluation) or the extraction of the pair amplitudes $\{t^{cc}_{kk}\} $ from $\{t^{cd}_{kl}\}$ (during energy calculation).
Our calculations suggest that these operations (data manipulations, intermediate creation) become a significant bottleneck on the CPU side for large numbers of basis functions.
Whether they can be efficiently offloaded to the CPU requires, however, further evaluation.
Thus, for large enough basis set sizes, we observe another shift in the bottleneck operations, moving from tensor contractions to data preparation.
Finally, we should stress that a direct comparison to CPU-only data is difficult primarily because the CPU-only counterpart implementations are computationally too time-consuming and exceed the available computing time on the underlying HPC infrastructure.

\section{Conclusions and outlook}\label{sec:conclusions}

In this work, we introduced new batching algorithms, namely, the asymmetric and dynamic C-split protocol and a fully generic splitting algorithm, to efficiently offload the bottleneck CCSD contraction to the GPU in the PyBEST software package (cf. Figure~\ref{fig:pybest_c-split}).
That, in combination with the NVIDIA H100 and GH200 GPU architectures, allows us to obtain a significant speedup for both the CCSD bottleneck contraction and CCSD vector function evaluation compared to our previous implementations.~\cite{pybest-gpu-jctc-2024}
Specifically, combining the C-split algorithm with PyTorch yields speedups of up to a factor of 10.

The C-split batching algorithm was applied to a molecular test set consisting of \ce{(H2O)10}, \ce{(mU)2 \cdot H2O}, and L0 (Figure~\ref{fig:structures}).
For these systems, we observe a substantial reduction in the CCSD iteration time.
For small to medium systems (up to 500 basis functions), CuPy on the NVIDIA GH200 provides the best performance, yielding overall speedups of 4–5 for the CCSD iteration step.
For these systems, PyTorch on H100 is consistently faster than CuPy, whereas CuPy performs slightly better on GH200.
For L0 in the aug-cc-pVDZ basis set (742 basis functions), PyTorch becomes slightly faster than CuPy on GH200.
For the largest system investigated---L0 in the cc-pVTZ basis set (exceeding 1000 basis functions)—only CuPy timings are presently available. Here, the CCSD iteration time on the GH200 is reduced by approximately 60\% relative to H100, consistent with the GH200's superior specifications, including higher peak memory bandwidth, larger on-GPU HBM capacity, and the high-bandwidth coherent NVLink-C2C interconnect to unified CPU memory.

Following targeted optimizations of the scaling-determining tensor contraction in the CCSD procedure within PyBEST, secondary operations (e.g., \texttt{numpy.einsum}-restricted operations, residual builds, data preparation) now contribute more significantly to the total runtime.
Further performance gains will therefore require optimization of these supporting kernels.
These benchmarks demonstrate that the choice between CuPy and PyTorch for maximal CCSD acceleration in PyBEST, specifically, or Python, in general, is not trivial: optimal backend selection depends sensitively on system size, basis set size, contraction pattern, and hardware platform (H100 vs. GH200).
To systematically address this complexity and automate optimal kernel dispatch, we plan to integrate machine learning techniques—such as performance modeling and heuristic-based selection—to predict and select the most efficient backend and implementation path for each contraction type and hardware configuration in PyBEST.
Future development will focus on leveraging multi-GPU parallelism on both the NVIDIA H100 and GH200 architectures, including exploitation of NVLink domain-spanning capabilities on GH200 systems to potentially enable larger-scale coupled-cluster calculations on systems with thousands of basis functions.
\section*{Conflicts of interest}
There are no conflicts to declare.
\begin{acknowledgement}
Funded/Co-funded by the European Union (ERC, DRESSED-pCCD, 101077420).
Views and opinions expressed are, however, those of the author(s) only and do not necessarily reflect those of the European Union or the European Research Council. Neither the European Union nor the granting authority can be held responsible for them. 

J.Ś., S.A., and P.T.~acknowledge financial support from the SONATA BIS research grant from the National Science Centre, Poland (Grant No. 2021/42/E/ST4/00302).

This work was completed in part at the Poland Open Hackathon, part of the Open Hackathons program. 
The authors acknowledge OpenACC-Standard.org for their support.
We gratefully acknowledge Polish high-performance computing infrastructure PLGrid (HPC Centers: ACK Cyfronet AGH and WCSS) for providing computer facilities and support within computational grant no. PLG/2025/018840.

\end{acknowledgement}

\section*{Data Availability Statements}
The data underlying this study are available in the published article.
The PyBEST code is available on Zenodo at \url{https://zenodo.org/records/10069179} and on PyPI at \url{https://pypi.org/project/pybest/}.
\bibliography{p}
\end{document}